\documentclass{sf2a-conf2023}
\usepackage{graphicx}
\usepackage{hyperref}
\usepackage[]{natbib}  
\usepackage{epstopdf}
\usepackage[perpage]{footmisc}

\def\BibTeX{{\rm B\kern-.05em{\sc i\kern-.025em b}\kern-.08em
    T\kern-.1667em\lower.7ex\hbox{E}\kern-.125emX}}
\bibpunct{(}{)}{;}{a}{}{,}  

\begin{document}

\TitreGlobal{SF2A 2023}


\title{A brief review on Fast Radio Bursts}

\runningtitle{Fast Radio Bursts}

\author{C. NG}\address{Laboratoire de Physique et Chimie de l'Environnement et de l'Espace - Université d'Orléans/CNRS, 45071, Orléans Cedex 02, France}

\setcounter{page}{1}


\maketitle


\begin{abstract}
This is a brief, non-exhaustive review of Fast Radio Burst (FRB), a new category of radio transients originating from extragalactic distances. We discuss the key observational properties known so far and the scientific applications of FRBs. We summarize the FRB-related research in the French astrophysics community, and conclude by sharing some insights to the future of FRB science. 
\end{abstract}

\begin{keywords}
Fast Radio Bursts, radio, transients
\end{keywords}


\section{Introduction}
Fast Radio Burst (FRB) is a recent astrophysical phenomenon, involving bright bursts ($<$10$^{44}$\,erg\,s$^{-1}$) coming from extragalactic distances, currently observed in the redshift range from $z\sim0.03$ to $z>1$.
They are extremely short in duration with bursts observed to last anything from tens of microseconds to several milliseconds, a fraction of the time it takes to blink an eye.
The inferred event rate of FRB is high, of the order of 5000 events per sky per day above a fluence of 2\,Jy\,ms \citep{Champion2016,Keane2015}, in other words, there is an FRB every few tens of seconds. 
The fact that FRBs are prolific, bright and extragalactic brings high hopes that we can use FRBs as a cosmological probe,
for example, to study the intergalactic medium \citep{ravi2016} and find the missing baryons \citep{Macquart2020}. 

The first FRB detection was made in 2007 \citep{Lorimer2007}, in the analysis of archival pulsar data taken in 2001. 
In the decade following, our knowledge of FRBs has
been limited since only a few tens of cataclysmic FRBs were observed, 
constrained by the field-of-view (FOV) of single dish radio telescopes (e.g. Parkes, Arecibo) used for FRB studies at the time.
Even though these are sensitive instruments, they can only see one point (or a few multi-beam positions) on the sky at a time. 
Since we did not know \textit{a priori} when or where FRBs burst, plus the fact that in the early days all of them seemed to be cataclysmic, one-off, non-repeating bursts, it has been really difficult to study them.

Since 2018, the number of FRB discoveries has exploded thanks to the CHIME radio telescope in Canada \citep{CHIME}. 
With the original design of a cylindrical transit telescope, 
CHIME has a FOV hundred times larger and this has proved to be a game changer for the FRB field, enabling hundreds of FRBs discoveries in the first years of the operation \citep{CHIMEcatalog1}.
However, while CHIME is undoubtedly a great survey instrument, it is not 
capable of localizing precisely where these FRBs come from. 
The localization uncertainty of CHIME is roughly 10" for the nearest one percent of sources it sees \citep[e.g.][]{Bhardwaj2021}, but the region can be as large as 1\,deg if the FRB was detected in the sidelobe \citep[e.g.][]{SGR-CHIME}. 
Typically there are dozens of galaxies in a field of this size and so we cannot know exactly from which one did the FRB originate.
To conduct proper host galaxy identification, we will need arcseond-level of spatial resolution provided by interferometer telescopes.
A number of interferometers have recently joined the FRB field, including the Deep Synoptic Array (DSA) in the US and the Australian Square Kilometre Array Pathfinder (ASKAP) in Australia. 
These facilities have been responsible for the increase in localized FRBs in the last few years. 
To date, according to the Transient Name Server (TNS) database\footnote{TNS database search engine: \url{https://www.wis-tns.org/search}} there are more than 750 FRBs published, of which almost 40 have a host localization, in additional to a lot of more FRBs yet to be officially announced. 

\section{Key observational properties of FRBs}
While most of the FRBs are one-off events, a small percentage of them have been seen to burst repeatedly. At the moment of writing, there are 51 published repeating FRBs, while a lot more are yet to be published. In \citet{CHIMErepeater3}, the burst cadence of 46 repeating FRBs are compared and it appears that they tend to cluster in time and energy distribution. The repeaters either go into an active state where it emits brighter and more bursts or they stay quiet. 
FRB~20180916B is a particularly active FRB.
Thanks to its high burst rate, a periodic activity window of about 16$\pm$0.15 days ($\sim$4 days active followed by 12 days inactive) has been detected \citep{CHIME2020}. 
Periodic activity window has been established for one other repeating FRB~121102 at about 160 days \citep{Rajwade2020,Cruces2021}. 
The first CHIME/FRB catalog \citep{CHIMEcatalog1} consists of over 500 FRBs and it is the first large sample of FRBs detected under the same system thus representing a good starting point for population analysis. 
Using that, \citet{Pleunis2021} reported statistically significant morphological differences between the repeaters and non-repeaters.  
Repeater bursts are intrinsically broader in width and narrower in bandwidth.
However, it is not clear if the one-off FRBs are truly cataclysmic events, or if we just have not observed them when they repeat, or perhaps the repeated bursts are below the detection threshold of the telescope used. 

The progenitor of FRB is another outstanding question. So far only one FRB has been unambiguously associated with the magnetar SGR~1935+2154 in our Milky Way \citep{SGR-CHIME,Bochenek2020}.
In 2020, SGR~1935+2154 underwent a period of X-ray burst storm which lasted for hours. The high energy activity was detected by a number of X-ray (NICER, Chandra, XMM, Swift XRT) and Gamma ray (Swift BAT, Fermi, NuSTAR, Integral ) telescopes \citep{Mereghetti2020}. FRB-like bursts were detected by two radio telescopes, namely CHIME and STARE2.
The X-ray peaks lag the radio ones by 6.5\,ms.
Magnetar has always been one of the popular models of the FRB phenomenon, and this discovery adds a clear vote to that theory. 
However, as mentioned earlier, there appears to be multiple populations of FRBs. It would be hard to explain all the observed FRB properties, for example the periodic activity seen in FRBs~121102 and 20180916B, with the magnetar flare model. 
This FRB is also Galactic instead of extragalactic, and quite a bit less luminous than the bulk of the FRBs detected so far. 
Is it possible that SGR~1935+2154 is an exception rather than the rule? The answer is not conclusive.
For all the other FRBs, no afterglow has been detected that can allow a progenitor association.

Nonetheless, almost 40 FRBs have been associated with a host galaxy. 
Most FRB hosts are star-forming, spiral galaxies, but there are also FRBs from more lenticular hosts. 
So far all hosts are located below z$\approx$1, but this is most certainly sensitivity limited and as better telescope facilities come online we will be able to see further.
\citet{Gordon2023} studied the overall properties of 23 highly-secure host galaxies of FRBs, involving six repeaters and 17 apparent non-repeaters. 
They find that FRB hosts have a median stellar mass of 
$\approx$10$^{9.9}\,$M$_{\odot}$, a mass-weighted age of $\approx$5.1\,Gyr, and an ongoing star formation rate of 
$\approx$1.3\,M$_{\odot}$\,yr$^{-1}$, while also noting a wide range in all properties.
They report no statistically significant distinction between the hosts of repeaters and non-repeaters.
We have to be careful when reading into this kind of analysis at a relatively early stage where the sample size remains modest. In fact, \citet{Mannings2021} presented examples of FRBs that are not located in the centre of their host galaxies but can, for instance, but on the spiral arms. That means an FRB associated with a star-forming galaxy does not necessarily imply that the FRB comes from the star-forming region. 
For example, FRB~20200120E in the M81 (star-forming galaxy) has been pin-pointed to a Globular cluster where there is little star forming \citep{Kirsten2022}. As an aside, 
a Globular cluster FRB makes a rather difficult case to reconcile with the magnetar model. 
In general, it seems that FRBs live in diverse local host environments and can come from various types of host galaxies. 

\section{Scientific applications of FRBs}
FRBs are the shortest-duration extragalactic transients, and the most compact known extragalactic sources of electromagnetic radiation. This means we can potentially use FRB as a cosmological probe. 

The dispersion measure of FRB (DM, not to be confused with Dark Matter) is arguably the most important observable of FRB studies, and it is a proxy for the distance of the FRB. The observed DM (DM$_{\mathrm{obs}}$) includes the total line-of-sight electron-column densities, and can contain contributions from the host-galaxy interstellar medium (ISM), the circumgalactic medium (CGM), the intracluster medium (ICM) of intervening systems,  the intergalactic medium (IGM), and the galactic halo of our Milky Way.
FRBs act as a back-light and provide a clean signal for the study of these components which are very hard to probe otherwise, with no need for additional simulations.

\citet{Ravi2019Fast} suggests that, with 1000 to 10,000 well localized FRBs, 
we would be able to make statistical detection and study the CGM, and understand whether all dark matter is composed of compact objects. 
There is already a study that has resolved the missing baryon problem \citep{Macquart2020}, and a recent detection of FRBs in the Abell cluster \citep{Connor2023} could allow for the study of the ICM. 
With 10,000 to 100,000 FRBs, we might be able to measure the mean IGM density and study the magnetic field in the cosmic web, achieve a global detection of He-II deionization, and detect baryon acoustic oscillations in DM-space clustering. 
Finally, with 100,000 to 1 million FRBs, we might be able to improve the kinetic Sunyaev-Zel'dovich (kSZ) constraints on large-scale structure growth rate.
Also, FRBs might provide the only means of measuring sub-ms time delays imparted by extragalactic lensing phenomena \citep{Sammons2020}. 

\begin{figure}[ht!]
 \centering
 \includegraphics[width=0.8\textwidth,clip]{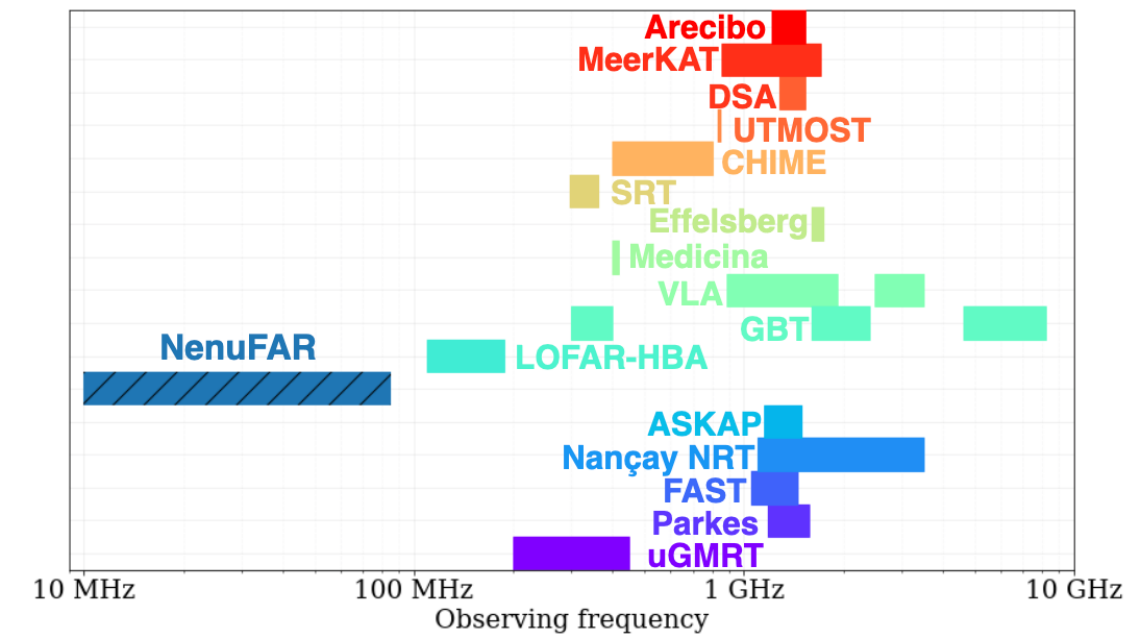}
  \caption{A non-exhaustive list of radio telescope facilities that have detected FRB burst to-date, spanning as low as 110\,MHz with the LOFAR-HBA to as high as 8\,GHz with the Green Bank Telescope (GBT). We have an on-going FRB monitoring campaign with NenuFAR, which operates in a much lower frequency range of 10--85\,MHz. Any future positive detections from NenuFAR will provide important clues regarding the emission mechanisms of FRBs.}
  \label{fig:Freq}
\end{figure}

\section{The French FRB community}
With these exciting prospects of FRB studies, we have in France a growing community, currently involving more than a dozen people across at least seven institutes. 
The largest focus of FRB work in France is in the domain of observations, led by researchers at the ORN\footnote{ORN: Observatoire Radioastronomique de Nançay}, LPC2E\footnote{LPC2E: le Laboratoire de Physique et Chimie de l’Environnement et de l’Espace}, SUBATECH\footnote{SUBATECH: le laboratoire de Physique SUBAtomique et TECHnologies associées} and LESIA\footnote{LESIA: le Laboratoire d’études spatiales et d’instrumentation en astrophysique}. 
Notably, there is an on-going campaign with the Nançay Extension Upgrading LOFAR (NenuFAR) radio telescope to follow up repeating FRBs \citep{Decoene2023}.
As shown in Fig.~\ref{fig:Freq}, FRBs have so far only been detected from 110\,MHz \citep{Pleunis2021lofar} up to 8\,GHz \citep{Zhang2018}.
With NenuFAR, we hope to open up a new window of FRB observations at the very low frequency range (10$-$85\,MHz), which is important for understanding the emission mechanisms of FRBs. 
The NenuFAR FRB project currently monitors 15 sources with at least 500 observations taken and analysis on-going. 
It has been found that FRB emission at low frequencies can be systematically delayed by a few days after the higher frequency detections, at least in the case of FRB~20180916B between its CHIME and LOFAR signals \citep{Pleunis2021lofar}. This delay is perfect for triggering NenuFAR observations as soon as other telescopes at higher observing frequencies send out alerts of burst activities. Triggering capability is currently not available on NenuFAR but could have a relevant use case for FRBs and other transient studies.
On the Nançay radio telescope (NRT), the Extragalactic Coherent Light from Astrophysical Transients (ECLAT) campaign has been monitoring 11 repeating FRBs since 2022.
Thanks to the high sensitivity and time resolution of the NRT, seven FRBs have been detected. 
In one of the sources, dense forests of clustered microshots in the bursts have been observed. They can be extremely bright, occasionally exceeding a signal-to-noise of 1000 \citep{Hewitt2023}.
These components have different dispersion measures, and they could be different burst types arising from different emission mechanisms, similar to what is seen in solar radio bursts. 
French astronomers are also involved in other international radio collaboration, for example, CHIME\footnote{CHIME: \url{https://chime-experiment.ca/en}} and MeerTRAP\footnote{MeerTRAP: \url{https://www.meertrap.org}}. In additional to radio, there are efforts at IRAP\footnote{IRAP: the Institut de Recherche en Astrophysique et Planétologie} to study potential X-ray emission of FRBs, as well as neutrinos work at SUBATECH. 
There is also multi-wavelength follow-up work for example the Deeper Wider Faster (DWF)\footnote{DWF:\url{https://www.swinburne.edu.au/research/centres-groups-clinics/centre-for-astrophysics-supercomputing/our-research/data-intensive-astronomy-software-instrumentation/deeper-wider-faster-program/}} 
campaign being carried out at CEA\footnote{CEA: Le Commissariat à l'\'{e}nergie atomique et aux \'{e}nergies alternatives}.
Finally, Machine Learning (ML) algorithms for FRB detections are being developed at LESIA and LPC2E. 
A number of theoretical studies of FRBs have also been presented by researchers from LUTH\footnote{LUTH: le Laboratoire Univers et Théories} and LPC2E \citep[see, e.g.][]{Bonetti2016,Bentum2017,Mottez2020,Decoene2021theory,Voisin2023}.

\section{Conclusion}
Looking to the future of FRBs, I am optimistic that we will soon uncover the origin of FRBs, both in terms of their provenance and the physics of the emission mechanisms of these energetic bursts. 
I think the most important clues will be obtained from Multi-wavelength FRB observations \cite[see, e.g. a review by][]{Ng2023}. 
We will soon have a large sample of well localized FRBs. 
Upcoming next-generation radio telescope facilities including CHORD \citep{CHORD}, DSA-2000 \citep{DSA2000} and the Square Kilometre Array (SKA) will provide over 500 mas-localized FRBs per month. 
A critical challenge associated with this is how will the identification of host galaxies be able to keep up with this discovery rate? 
Finally, there is high hopes that we can use FRBs as cosmological probes for our Universe, for the studies of the missing baryons, galactic halo, setting constraints on Hubble constant (H$_{0}$), He-II reionization, lensing, IGM magnetic field, to name but a few.  
In summary, FRB is a young and fast growing field with a bright future and lots more discoveries to come. 

\bibliographystyle{aa}  
\bibliography{NG-GUIHENEUF1} 

\end{document}